\documentclass[nofootinbib,twocolumn]{revtex4-2}
\usepackage[T1]{fontenc}
\usepackage{graphicx}
\usepackage{rotating}
\usepackage{bm}        
\usepackage{amssymb}   
\usepackage{bm}
\usepackage{float}
\usepackage{tikz}
\usepackage{diagbox}
\usepackage{braket}
\usepackage{color} 
\usepackage{colortbl}
\usepackage{amsmath}
\usepackage{xcolor}
\usepackage{soul}
\usepackage{float,subcaption}
\usepackage{physics}

\makeatletter

    \def\CT@@do@color{%
      \global\let\CT@do@color\relax
            \@tempdima\wd\z@
            \advance\@tempdima\@tempdimb
            \advance\@tempdima\@tempdimc
    \advance\@tempdimb\tabcolsep
    \advance\@tempdimc\tabcolsep
    \advance\@tempdima2\tabcolsep
            \kern-\@tempdimb
            \leaders\vrule
                    \hskip\@tempdima\@plus  1fill
            \kern-\@tempdimc
            \hskip-\wd\z@ \@plus -1fill }
    \makeatother

\def\k1{k_1}
\def\k2{k_2}
\def\q1{q_1}
\def\q2{q_2}

\def\({\left (}
\def\){\right )}
\def\[{\left [}
\def\]{\right ]}

\newcommand{\beq}{\begin{equation}}
\newcommand{\eeq}{\end{equation}}

\DeclareMathAlphabet\mathbfcal{OMS}{cmsy}{b}{n}

\begin{document}


\title{Bayesian autotuning of Hubbard model quantum simulators}

\author{Ludmila Szulakowska, Jun Dai}
 \affiliation{
Department of Chemistry, University of British Columbia, Vancouver, B.C. V6T 1Z1, Canada \\
Stewart Blusson Quantum Matter Institute, Vancouver, B.C. V6T 1Z4, Canada }



\date{\today}

\begin{abstract}
Spins in gated semiconductor quantum dots (QDs) are a promising platform for Hubbard model simulation inaccessible to computation. Precise control of the tunnel couplings by tuning voltages on metallic gates is vital for a successful QD-based simulator. However, the number of tunable voltages and the complexity of the relationships between gate voltages and the parameters of the resulting Hubbard models quickly increase with the number of quantum dots. As a consequence, it is not known if and how a particular gate geometry yields a target Hubbard model. To solve this problem, we propose a hybrid machine-learning approach using  a combination of support vector machines (SVMs) and Bayesian optimization (BO) to identify combinations of voltages that realize a desired Hubbard model. SVM constrains the space of voltages by rejecting voltage combinations producing potentials unsuitable for tight-binding (TB) approximation. The target voltage combinations are then identified by BO in the constrained subdomain. For large QD arrays, we propose a scalable efficient iterative procedure using our SVM-BO approach, which optimises voltage subsets and utilises a two-QD SVM model for large systems. Our results use experimental gate lithography images and accurate integrals calculated with linear combinations of harmonic orbitals to train the machine learning algorithms. 
\end{abstract}

\maketitle


\section{Introduction}

    Recent progress in manipulating spins in gated semiconductor quantum dots (QDs) \cite{maune_coherent_2012, maurand_cmos_2016, hendrickx_four-qubit_2021} offers an opportunity for realizing scalable quantum systems for applications in quantum computing \cite{loss_quantum_1998,zwanenburg_silicon_2013,noauthor_two-qubit_nodate,altintas_spin-valley_2021,wang_ultrafast_2022,blumoff_fast_2022} and simulation, such as Kitaev chain or Hubbard model simulation \cite{byrnes_quantum_2008,  jaworowski_macroscopic_2017, hensgens_quantum_2017, dehollain_nagaoka_2020, dvir_realization_2022}. Hubbard models with many sites, inhomogeneous or time-varying parameters remain largely unexplored due to computational complexity, but can be studied experimentally with new generations of gated QD arrays.  A successful QD-based device appropriate for this goal must provide means to control the charge occupation, chemical potential of QDs and tunnel couplings with considerable precision in order to prepare, manipulate and read out many-particle quantum states \cite{loss_quantum_1998, levy_universal_2002,hanson_spins_2007,zwanenburg_silicon_2013}. Therefore, developing the tools to identify the charge states and tune their properties to desired operating regimes is vital for scaling up the complexity of quantum simulators and discovering new physical phenomena, such as topological phases \cite{mourik_signatures_2012,jaworowski_macroscopic_2017, chevallier_topological_2018, perez-gonzalez_simulation_2019, kiczynski_engineering_2022} and strongly-correlated ground states \cite{hanson_spins_2007, byrnes_quantum_2008, hensgens_quantum_2017, dehollain_nagaoka_2020, dvir_realization_2022, saleem_quantum_2022}. In particular, the inter-dot tunneling amplitude requires special attention as it determines the exchange interaction of spins in QDs \cite{diVincenzo_universal_2000, hanson_spins_2007}, which affects all their applications, ranging from qubit design to parameters of interacting electron models under study. 
    
Design of QD-based simulator relies on well-established techniques of trapping electrons in electrostatic potential wells, i.e. QDs, created at the interface of semiconductor devices build with silicon \cite{lim_observation_2009,maune_coherent_2012, maurand_cmos_2016}, germanium \cite{kawakami_electrical_2014,hendrickx_four-qubit_2021}, III-V materials \cite{ciorga_addition_2000,gaudreau_stability_2006,mar_bias-controlled_2011,dehollain_nagaoka_2020} or 2D semiconductors \cite{song_gate_2015, zhang_electrotunable_2017, pisoni_gate-tunable_2018, bieniek_effect_2020, szulakowska_valley-_2020, boddison-chouinard_gate-controlled_2021, altintas_spin-valley_2021, jing_gate-controlled_2022}. This confining potential landscape is determined by a set of voltages applied to lithographically fabricated metallic gates placed on top of the nanomaterial. Contacts are reservoirs of electrons placed at the edges of the structure to allow for electron tunneling into the confining potential wells, where they can be manipulated. Barrier gates are used to control tunneling between QDs, while plunger gates are designed to alter the depth of each potential well. Changing the voltages on all those gates allows for realizing a vast range of electron states for various applications. Characterization of charge states achieved with different voltage combinations is usually performed by repeatedly measuring the charge stability diagram – an image of transport features as a function of gate voltages, essential for experimentally tuning the system to a desired regime \cite{gaudreau_stability_2006,wang_quantum_2011}.

This approach to control the gate voltages limits the scalability of QD-based simulators. For systems with many quantum dots, the number of gates is large and the design complexity makes the voltage calibration process impractical for manual tuning. Moreover, in dense devices, the relative proximity of gates produces substantial cross-talk, which further complicates the independent QD control \cite{baart_computer-automated_2016, hensgens_quantum_2017, van_Diepen_automated_2018,teske_machine_2019}. An additional obstacle for practical QD simulators is the presence of charge impurities unavoidable in the fabrication process, which alter the potential landscape and lead to non-uniform device performance \cite{baart_computer-automated_2016, hensgens_quantum_2017, van_Diepen_automated_2018}. These challenges, combined with variations of gate geometry, and a wide range of material parameters and screening effects impede the development of practical tools for experimental control of QD-based simulators.

Machine learning (ML) has emerged as a promising tool for some of the experimental challenges with QD control  \cite{van_Diepen_automated_2018}. Deep neural networks \cite{turaga_convolutional_2010, lecun_deep_2015,kalantre_machine_2019, durrer_automated_2020, zwolak_autotuning_2020, darulova_evaluation_2021, oakes_automatic_2021, schuff_identifying_2022}, image recognition \cite{lecun_handwritten_1989, lecun_deep_2015, mills_computer-automated_2019,  teske_machine_2019, durrer_automated_2020, lapointe-major_algorithm_2020, krause_learning_2022} and supervised classification \cite{lecun_deep_2015, darulova_autonomous_2020, schuff_identifying_2022} have been demonstrated to aid charge state characterization \cite{kalantre_machine_2019, mills_computer-automated_2019, darulova_autonomous_2020, durrer_automated_2020, lapointe-major_algorithm_2020}, coupling parameter tuning \cite{van_Diepen_automated_2018} and gate voltage optimization \cite{baart_computer-automated_2016, kalantre_machine_2019, moon_machine_2020, lapointe-major_algorithm_2020, zwolak_autotuning_2020, schuff_identifying_2022} in a single QD \cite{darulova_autonomous_2020, lapointe-major_algorithm_2020, zwolak_autotuning_2020}, double QDs \cite{baart_computer-automated_2016, van_Diepen_automated_2018,  teske_machine_2019, darulova_autonomous_2020,durrer_automated_2020, zwolak_autotuning_2020, darulova_evaluation_2021}, triple QDs and arrays of QDs \cite{baart_computer-automated_2016, van_Diepen_automated_2018, kalantre_machine_2019, mills_computer-automated_2019, oakes_automatic_2021}. Unsupervised statistical methods \cite{moon_machine_2020, lennon_efficiently_2019} and deterministic algorithms \cite{baart_computer-automated_2016, darulova_autonomous_2020, lapointe-major_algorithm_2020, krause_learning_2022} have also been used for double-QD tuning. ML also proven useful for compensating for cross-capacitance in devices \cite{oakes_automatic_2021}, calibration of virtual gates in place of real ones \cite{mills_computer-automated_2019, oakes_automatic_2021} and in the analysis and parameter-extraction from charge stability diagrams \cite{mills_computer-automated_2019}. 

A vast majority of these approaches relied on experimentally obtained data as input \cite{mills_computer-automated_2019,teske_machine_2019} or intermediate step in a feedback protocol \cite{baart_computer-automated_2016, van_Diepen_automated_2018,darulova_autonomous_2020,  durrer_automated_2020, lapointe-major_algorithm_2020, moon_machine_2020}, which required numerous measurements or involved readjustments and recapturing procedures \cite{baart_computer-automated_2016, van_Diepen_automated_2018, darulova_autonomous_2020, durrer_automated_2020, lapointe-major_algorithm_2020, zwolak_autotuning_2020, lennon_efficiently_2019, moon_machine_2020}. Although scarcity of experimental data has been addressed in Refs \cite{zwolak_autotuning_2020, darulova_evaluation_2021} with synthetic data, many ML solutions for QD simulators suffer from crude theoretical assumptions. This includes the Thomas-Fermi approximation for electron density \cite{kohn_nobel_1999, darulova_evaluation_2021}, the use of exponential fits to tunneling couplings \cite{mills_computer-automated_2019} or constant interaction model with weak coupling and absent barrier gates \cite{krause_learning_2022}, which limits their applicability to a wider range of designs and materials. Another limitation of the optimization techniques used in Refs \cite{teske_machine_2019,  oakes_automatic_2021} is the need for obtaining the gradients of gate voltages in the parameter search, which may be prone to vanishing gradient problem \cite{kolen_gradient_2009}. The relationship between gate voltages and the parameters of the resulting quantum simulator is further complicated by the scarcity of the physical simulation domain: the majority of the voltage combinations produce unphysical potentials. Thus, an automated first-principle design of quantum simulators must be able to recognize the physical subdomain of experimentally tunable parameters.

To address this problem, we propose a hybrid machine-learning approach using  a combination of support vector machines (SVMs) and Bayesian optimization to identify combinations of voltages that realize a desired Hubbard model. SVM constrains the space of voltages by rejecting voltage combinations producing potentials unsuitable for tight-binding (TB) approximation. The target voltage combinations are then identified by Bayesian optimization (BO) in the constrained subdomain. We perform BO of gate voltages to produce a double QD system with tailored tunneling parameter $t$ and on-site Hubbard energy $U$, using experimental gate lithography images as input for realistic calculations of $t$ and $U$ with the linear combination of harmonic orbitals method (LCHO) \cite{puerto_gimenez_linear_2007}. This approach allows us to predict $t$ and $U$ for variable electrode design and with flexible material parameters and custom heterostructures. Our BO procedure operates without gradients or input from experimental charge stability diagrams, which are tedious to measure for large systems. BO is also suitable for problems with multiple local optima and noisy data. We also develop an iterative, scalable SVM-BO approach for multiple-site arrays, which is able to reach optimal solution by only optimising subsets of voltages at a time and uses only the two-site SVM. We predict the optimal voltage combination needed in experiment to prepare an on-demand double QD Hubbard model within 1.5\% error for model gates and 6\% for experimental gates, as well as for three-site system with $10\%$ error. This procedure can be combined with existing methods of preparing a charged state within quantum dots \cite{baart_computer-automated_2016, kalantre_machine_2019, mills_computer-automated_2019, darulova_autonomous_2020, durrer_automated_2020, lapointe-major_algorithm_2020,  moon_machine_2020, zwolak_autotuning_2020, schuff_identifying_2022}.

This paper is organized as follows: Section I describes the numerical calculations of the electrostatic potentials from the metallic gate input image by solving Poisson’s equation. Section II presents the SVM model developed to classify voltage combinations and demonstrates its excellent performance for rejecting undesired voltage combinations for two gate designs: an ideal simple-shaped gate set and a realistic experimental gate set obtained from lithography images. Section III describes the LCHO method for the calculation of the Hubbard model parameters  and presents the results for possible $U/t$ ratios that can be achieved with various gate designs. The adaptation of BO for the title problem is described in Section IV.  We present our results on optimal voltage combinations for the model and experimental sets of gates in Section V. 

    \section{Electrostatic potential from metallic gates \label{poissonSec}}

\begin{figure}[H]
    \centering
    \includegraphics[width=1\linewidth]{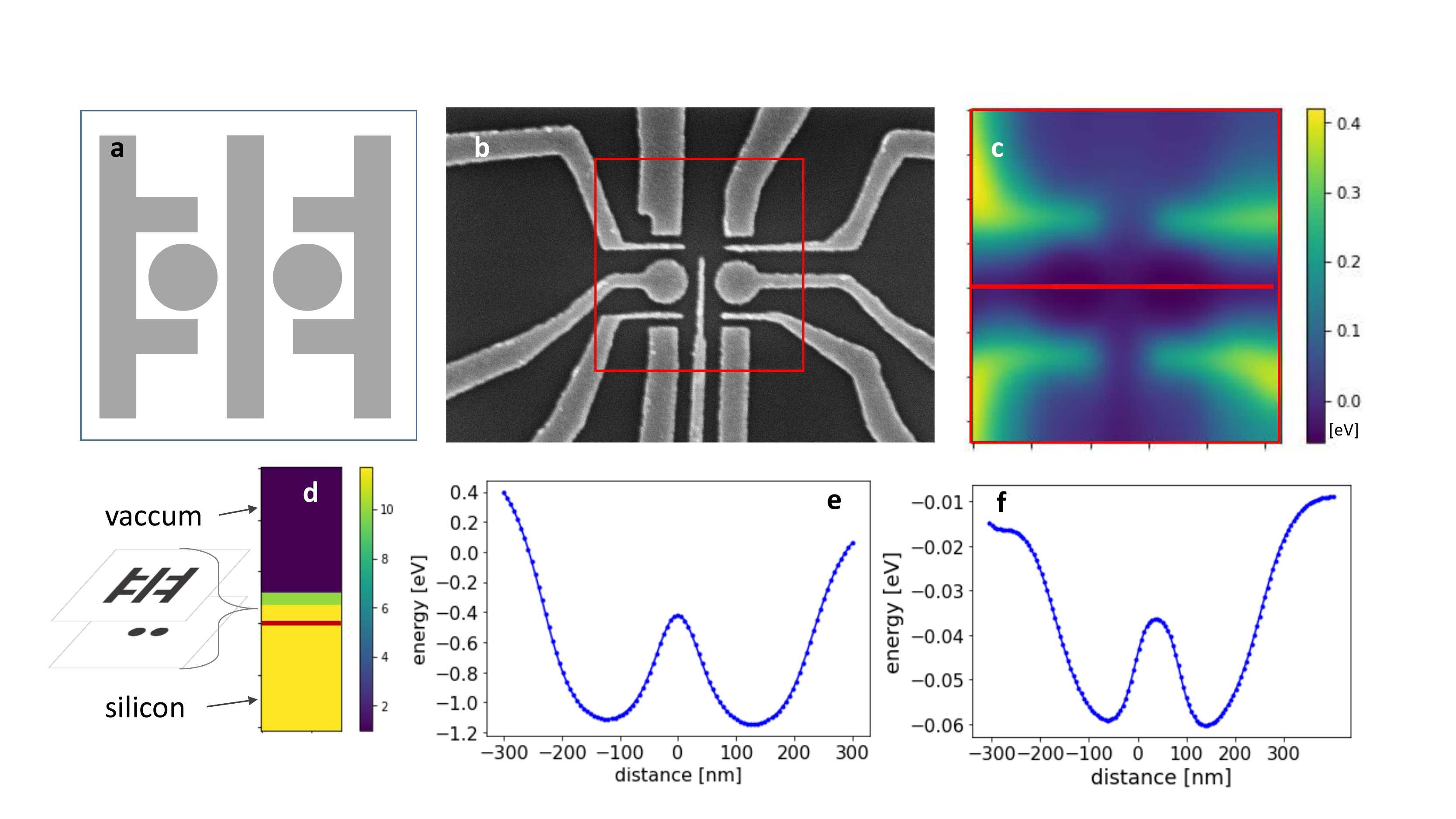}
    \caption{a) Model gate design with 2 confining plunger gates and 3 barrier gates controlling the tunnelling. b) Image of experimental gate design with 2 plunger gates, 5 barrier gates and 4 accumulation gates. Red box corresponds to the 2D potential landscape shown in c). Two potential wells are visible (dark blue) in c) Red line corresponds to crosssection potential in f). d) Semiconductor heterostructure design crosssection along z axis used for numerical Poisson equation solution. Metallic gates are placed inside the heterostructure with silicon substrate. Colors denote dielectric constant regions. Red line corresponds to 2DEG subject to confinement. e) Confining 2QD potential for model gate deisgn (cut along x). f) Confining 1QD potential for experimental gate design (cut along x).}
    \label{fig1}
\end{figure}

We begin by computing the electrostatic confining potential produced by two sets of metallic gates: the model basic-shape gates in Fig. \ref{fig1} and a realistic gate image from experiment \cite{sajadi_2022} in Fig. \ref{fig1}. Both sets of metallic gates are assumed to be 10 nm tall and are placed within a heterostructure inside a material with $\epsilon=10$ (Fig. \ref{fig1}). The pattern of the gates serves as input to a finite difference numerical method solving the Poisson's equation:
\begin{equation}
\nabla\cdot\Big(\epsilon(\boldsymbol{r})\nabla V(\boldsymbol{r})\Big)=-\frac{\rho(\boldsymbol{r})}{\epsilon_0}, \label{poissonVarE}
\end{equation}
where $\boldsymbol{r}=(x,y,z)$ is the position vector in 3D space, $\rho$ is the charge density, $\epsilon(\boldsymbol{r})$ is the dielectric constant and $\epsilon_0$ is the permittivity of free space.

We use finite difference method with the varied dielectric constant and two types of boundary conditions: the Dirichlet boundary condition at the top and bottom of computational box as well as inside the box, where the gate is placed, the Neumann boundary condition at the sides of the computational box, in $x$ and $y$ direction. We use a grid of 150$\times$150$\times$127 points in $x\times y\times z$ directions. The sample is modeled with $30nm$ of vaccum above the heterostructure and $30nm$ wetting layer below the metallic gate layer. The successive over-relaxation technique \cite{Saad2003} is used to speed up the numerical procedure.

Fig \ref{fig1} shows the resulting 2D electrostatic potential (for experimental gates) for an sample set of voltages, consisting of two confining wells (quantum dots) which act as sites in a 2-site Hubbard model. The red line marks a 1D cut of the potential passing through the minima of the two wells. The 1D example potential cuts for model as well as experimental gates are shown in Fig. \ref{fig1}.

\section{Hubbard model simulation domain}

In a typical experiment with gated QDs, each QD is tuned by $3$ electrodes.      
As the number of QDs increases as needed for many-site Hubbard models, 
the space of gate voltages becomes high-dimensional. More importantly, the vast majority of the gate voltage combinations results in electrostatic potentials that are unsuitable for quantum simulation of Hubbard models and must, therefore, be discarded as unphysical. Identifying the target gate voltages thus amounts to optimization in a highly constrained subdomain.

In order to identify the subdomain of gate voltages producing suitable potentials, we develop an SVM filter of gate voltages. We use SVM to solve a classification problem as implemented in the \textit{scikit-learn} python library \cite{scikit-learn} with a radial basis function (RBF) kernel. The SVM models are trained by the results of single-particle (SP) quantum calculation. For a given combination of voltages, we obtain the electrostatic potential as described in Section \ref{poissonSec} and solve the Schr\"{o}diger equation to obtain 
the particle density $p_{i \in [1,N_0]}$ for $N\leq N_0$ lowest-energy eigenstates.

We adopt the following criteria for the classification problem. The gate voltages are accepted as suitable for quantum simulation of the Hubbard models provided:
a) a significant portion $p=p_1\cdot p_2>p_0$ of the particle density in well 1 or 2 ($p_1$ or $p_2$) is enclosed within a given radius $R$ from the centres of the quantum dots (and the charge density within any single well does not vanish); b) a chemical potential $\mu$ is set and all $N<N_0$ SP energy levels $E_i$ are populated, i.e. $E_{N_0}\leq\mu$. For the present calculations,  we use $N_0=6, p_0=0.05, R=50$ nm, $\mu=-0.15$ eV. SVM is trained over $5\cdot10^4$ training points and achieves approximately $99\%$ overall success and above $98\%$ rejection success.

\begin{figure}[H]
    \centering
    \includegraphics[width=\linewidth]{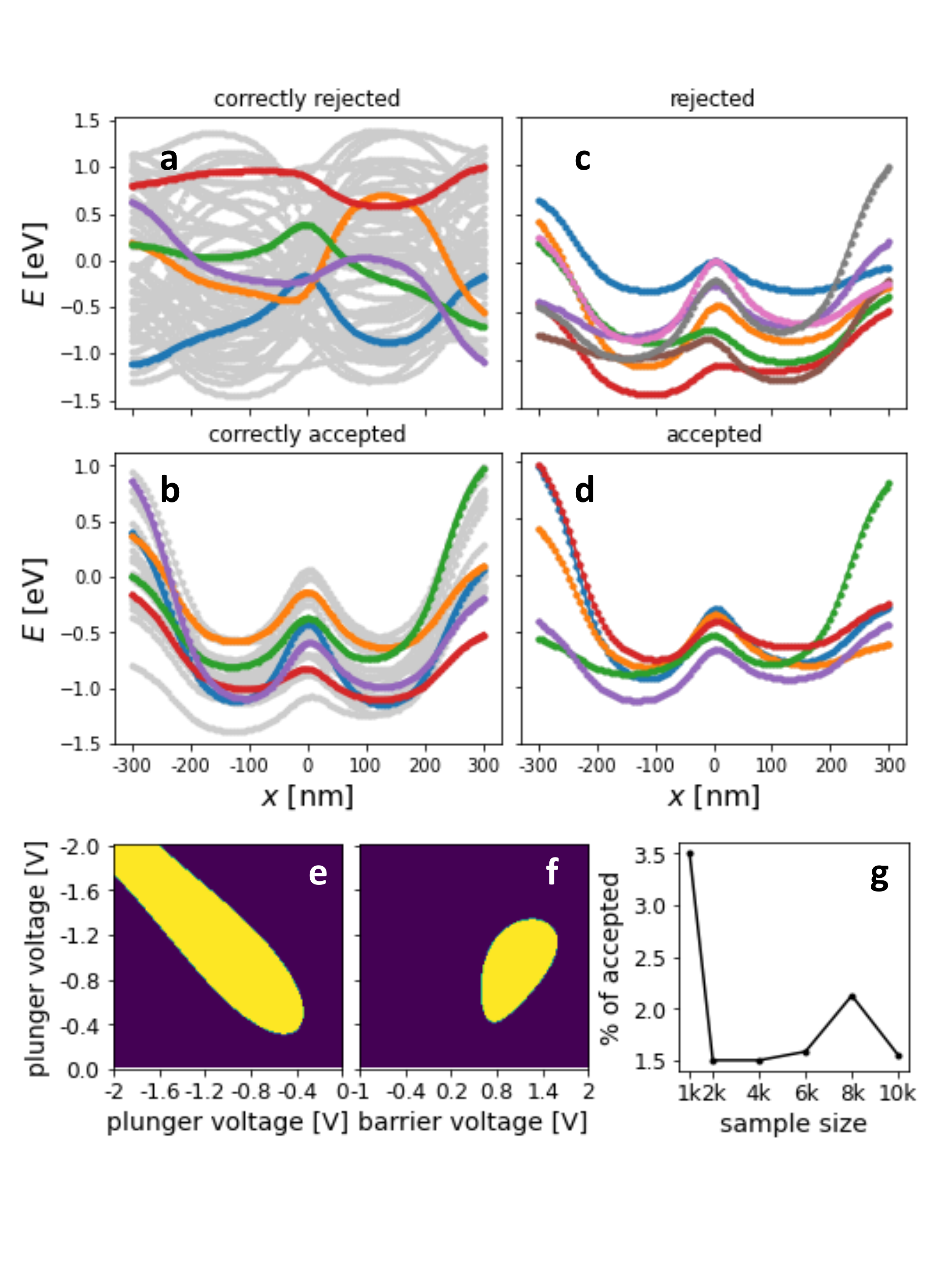}
    \caption{a) (b) Potentials classified as rejected (accepted) during SVM testing, in agreement with training labels. Grey lines represent all potentials included in classification. The potentials in b) follow a unified trend, while those in a) do not.  c) (d)  Potentials classified as rejected (accepted) during SVM testing, contrary to training labels. c) and d) are borderline examples and results depend on training label parameters. It is apparent that the SVM classification accepts a good selection of potentials for Hubbard model. e) (f) Classification of potentials in space of voltages for cuts along plunger-plunger (plunger-barrier) voltage plane. Percent of accepted potentials during test as a function of the size of the training sample (model gates). The accepted portion is of the order of 2\% for large enough sample size.}
    \label{fig2}
\end{figure}

The classification problem we consider here is imbalanced, i.e. around $2\%$ of the potentials are suitable for Hubbard model. In practice, it is important to tune this step to be sensitive to rejection in order not to perform further optimisation on unphysical potentials. Fig. \ref{fig2} a) and b) show examples of potentials from model gates which have been correctly accepted and rejected respectively. The grey line represent all potentials included in classification. It is apparent that the accepted potentials all exhibit an acceptable shape for TB, while rejected onces have chaotic shapes. Fig \ref{fig2} c) shows examples of incorrectly accepted potentials, which are still physical and can be confidently passed on to the next step. The results of Fig. \ref{fig2} show that the filter can be reliably used to respect the TB assumption and produce two-site potentials suitable for Hubbard model simulation. 

Fig \ref{fig2} d) shows example cuts within the high-dimensional voltage space with the classification label given by the present SVM model (yellow/ purple is accepted/rejected), which illustrates the significant reduction of the search space.

For more than two QDs, the space of voltages becomes even more restrictive and building a reliable SVM model is challenging. We found that approx. $0.01\%$ and approx. $10^{-4} \%$ of potentials were acceptable for three and four QDs, respectively. This is mainly because for more QDs, the first several SP energy shells should be aligned in order to allow tunnelling, and it is exponentially harder to achieve with growing number of voltages.  However, the primary aim of the SVM model in our optimisation is to identify the physical potentials for Hubbard model parameter calculation. If this task is done correctly, the alignment of energy levels can be achieved by a simple check following the SVM classification. Due to this, we find that it is sufficient to use the SVM model for two QDs to classify subsets of voltages for more QDs, provided that the gate geometry does not change significantly with more QDs. Detailed explanation of how this is used in practice is given in Sec. \ref{BOsection}.

\section{The Hubbard model parameters}

We seek to design and optimize the quantum simulator of a Hubbard model, where the quantum dots act as sites for charges. The Hamiltonian of the Hubbard model with $M$ spin-degenerate orbitals per dot can be written as follows:
\begin{equation}
\hat{H}_{H}=\sum_{\sigma,i,\lambda}e_{i\lambda}c_{i\sigma\lambda}^{\dagger}c_{i\sigma\lambda}+\sum_{\sigma ij\lambda\gamma}t_{ij\lambda\gamma}c_{i\sigma\lambda}^{\dagger}c_{j\sigma\gamma}+\sum_{i\lambda}U_{i\lambda}n_{i\lambda\downarrow}n_{i\lambda\uparrow}, \label{Hhub}
\end{equation}
where $i$ and $\lambda$ index orthogonal orbitals in individual sites, $c_{i\sigma\lambda}^{\dagger}(c_{i\sigma\lambda})$ is the creation (annihilation) operator of a particle on site $i$, orbital $\lambda$ with spin $\sigma$ and $e_{i\lambda},t_{ij\lambda\gamma}$ and $U_{i\lambda}$ are the parameters of the desired Hubbard model. 
  
The Hubbard model parameters are uniquely determined by the confining potential created by the  metallic gates and can be calculated for a given potential with the LCHO method \cite{puerto_gimenez_linear_2007}. In order to use LCHO, we consider the parts of the numerical potential that correspond to individual sites and solve the SP single-QD problem for each one of them. This serves as a basis for many-site LCHO calculation. In our optimisation procedure, the QD separation $a$ is varied approximately by translating the single-QD solution in space to calculate integrals at variable distances.

We now briefly describe the LCHO method. The dimensionless SP Hamiltonian in the two dimensional potential of a double QD reads:
\begin{equation}
\hat{H}_{SP}=-\frac{\partial}{\partial x^2}-\frac{\partial}{\partial y^2}+\sum_iV_i, \label{HSP}
\end{equation}
where we express all distances in effective Bohr radii $a_B^*=\epsilon\hbar^2/m^*e^2$ and all energies in units of the effective Rydberg $Ry^*=e^2/2\epsilon a_B^*$ and $m^*=0.06$, $\epsilon=16$ and $a_B^*=14.1$nm. In Eq. \ref{HSP} $V_i$ are the potentials for a single quantum dot $i$, which are well approximated by a 2D harmonic oscillator (HO) potential at the bottom of the well $V_j\approx -V_0\big( 1-\frac{(\boldsymbol{r}-\boldsymbol{R}_j)^2}{d^2}+\delta V\big)$, where $V_0$ denotes the strength of the potential on dot $j$, $\boldsymbol{R}_j$ is the centre position of the quantum dot, $d$ is the characteristic width and $\delta V$ represents deviation from the HO potential. This motivates the choice of the basis as the set of 2D HO eigenfunctions $\{\phi_{nm}^j(x,y)\}$ with corresponding eigenvalues $e_{nm}^j=-V_0+\omega_j(n+m+1)$. The 3D wavefunction of a particle is a product of the linear combination of HO orbitals on each dot 
\begin{equation}
\psi(x,y)=\sum_{inm}A_{nm}^i\phi_{nm}^i \label{LCHOwav}
\end{equation}
with the eigenfunction of the infinite narrow square quantum wall in $z$ direction $\xi(z)=\frac{1}{\sqrt{2}}\sin{k_zz}$, $\Psi(\boldsymbol{r})=\psi(x,y)\xi(z)$. Substituting Eq. \ref{LCHOwav} in the eigenvalue problem of Eq. \ref{HSP} and multiplying by $\phi_{j\beta}$ on the left we get
\begin{equation}
\sum_{i\alpha}\bra{\phi_{j\beta}}\hat{H}_{SP}\ket{\phi_{i\alpha}}A_{i\alpha}^l=E_l\sum_{i\alpha}\bra{\phi_{j\beta}}\ket{\phi_{i\alpha}}A_{i\alpha}^l, \label{HSPmat}
\end{equation}
where $\alpha$ denotes a composite index $\alpha=(n,m)$ and $l$ indexes eigenstates of Eq. \ref{HSP}. Eq. \ref{HSPmat} is the generalised eigenvalue problem for $\hat{H}_{SP}$: $H_{SP}A^l=E_lSA_l$, where $S$ is the overlap matrix with elements $\bra{\phi_{j\beta}}\ket{\phi_{i\alpha}}$. 

The parameters $e_{i\lambda}$ and $t_{ij\lambda\gamma},$ of the Hubbard model Hamiltonian in Eq. \ref{Hhub} can be found as the matrix elements of the Hamiltonian  matrix in the basis of orthogonal orbitals $\{\tilde{\phi}_{i\lambda}(x,y)\}$: $\tilde{H}=S^{-1/2}H_{SP}S^{-1/2}$.

\begin{figure}[H]
    \centering
    ~~~\includegraphics[trim=6cm 0 4.6cm 0,clip=true,width=1.09\linewidth]{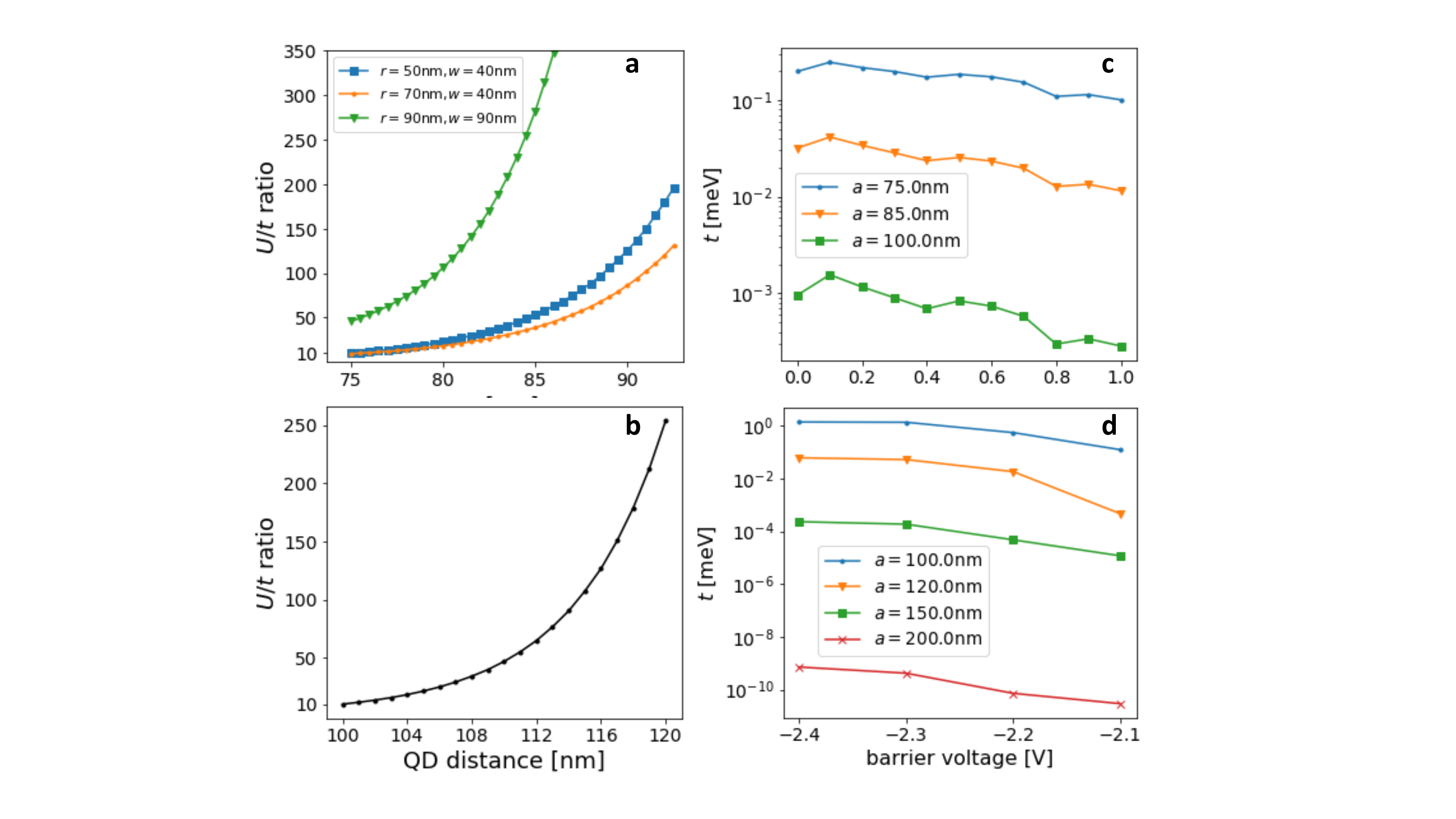}
    \caption{ a) (b) U/t ratios for model (experimental) design as a function of QD distance $a$. Variable dot radii $r$ and barrier widths $w$ are considered for model design. c) (d) Magnitude of tunneling integral $t$ as a function of middle barrier voltage for model (experimental) design for several QD distances $a$.}
    \label{fig3}
\end{figure}

The parameters $U_{i\lambda}$ from Eq. \ref{Hhub} can be obtained by evaluating the many-particle integrals in the orthogonal basis:
\begin{equation}
\begin{split}
V_{ijkl\alpha\beta\lambda\gamma}&=\bra{i\alpha; j\beta}\hat{V}_C\ket{k\lambda;l\gamma}\\
&=\int d\boldsymbol{\vec{r}}d\boldsymbol{\vec{r'}}\tilde{\phi}_{i\alpha}^*(\boldsymbol{\vec{r}})\tilde{\phi}_{j\beta}^*(\boldsymbol{\vec{r'}})\frac{2}{|\boldsymbol{\vec{r}}-\boldsymbol{\vec{r'}}|}\tilde{\phi}_{k\lambda}(\boldsymbol{\vec{r'}})\tilde{\phi}_{l\gamma}(\boldsymbol{\vec{r}}),
\end{split}
\end{equation}
where $\boldsymbol{\vec{r}}$ is a 3D vector, $V_C$ is Coulomb interaction and $U_{i\lambda}=\bra{i\lambda; i\lambda}\hat{V}_C\ket{i\lambda;i\lambda}$. We perform these 6D integrals using the \textit{vegas} phyton library \cite{lepage_adaptive_2021, vegas}, which allows for Monte Carlo estimates of arbitrary multidimensional integrals.

In Fig. \ref{fig3} we show examples of resulting Hubbard integrals for a two-site system, where $U=U_{i\lambda=(0,0)}$ and $t=t_{ij\lambda=\gamma=(0,0)}$. As expected, as the quantum dot distance decreases, the $U/t$ ratio drops (Fig \ref{fig3} a). Wider barriers decrease hopping and increase the $U/t$ ratio, while bigger QD radii cause the $U/t (a)$ dependence to grow at a smaller rate. Fig \ref{fig3} b) shows the value of $t$ for a selected voltage combination with the middle barrier voltage variable, for several quantum dot distances. As the barrier grows, the hopping is reduced.

    \section{Bayesian optimization of voltages \label{BOsection}}
    
    Here we present the algorithm we use to define the optimal set of gate voltages in order to realise a desired Hubbard model experimentally with a chain of electrostatically defined quantum dots. The Hubbard model is defined by the $t_{ij\lambda\gamma},U_{i\lambda}$ integrals given in Eq. \ref{Hhub}, and, for constant QD radius $R$ - mainly by $t_{ij\lambda\gamma}$, which becomes the optimised quantity in our procedure, depicted in Fig. \ref{fig4}. We set a goal for the value of $\tilde{t}$ value and vary the voltages $\{Z_i\}$ as well as the quantum dot distance $a$ iteratively until an optimal combination is found. Changing the set of voltages and distance $\boldsymbol{X}=(Z_0,Z_1,...,a)$ modifies the potential landscape, which is evaluated by the SVM classifier for suitability for Hubbard model realisation. Only the accepted potentials are used in the integral calculation stage to yield $t,U$. Finally, a new sample $\boldsymbol{X'}=(Z_0',Z_1',...,a')$ is selected and the classification and evaluation steps are repeated.

\begin{figure}[H]
    \centering
    \includegraphics[trim=0cm 0 0 0,width=0.9\linewidth]{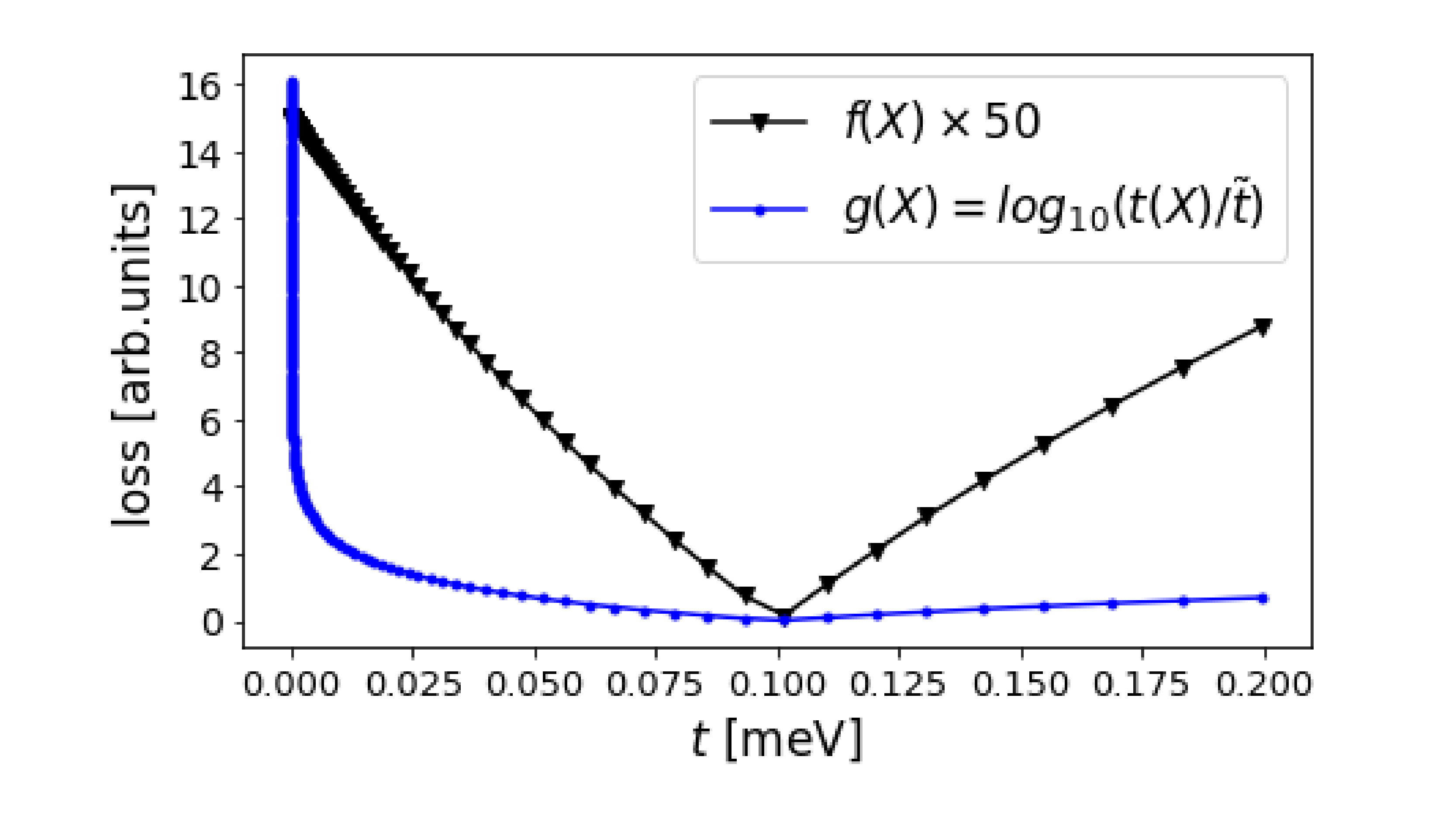}
    \caption{BO loss function used in this work (black traingles, magnified by a factor of 50) vs. function $g(\boldsymbol{X})$ (blue dots) defined as the logarithm of $t(\boldsymbol{X})/\tilde{t}$. The functions differ at low $t(\boldsymbol{X})$, where $g$ ($f$) varies fast (slowly) with the order of $t$.}
    \label{fig4a}
\end{figure}

We use Bayesian optimisation (BO) \cite{Mockus1978, snoek_practical_2012} to find the optimum voltage and distance combination $\boldsymbol{X}_{lim}$, which produces integrals as close to the desired  $\tilde{t}$ value as possible. We choose this method due to the black-box nature of the evaluated function and because no derivative calculation is needed. Also, as several combinations $\boldsymbol{X}$ may produce a desired Hubbard model, BO is particularly suitable because of its suitability for non-convex problems. 

For the two-site problem, with a constant size of plunger gates, $U$ integrals are slowly varying and achieving a variety of values would require size adjustment. For a costant diameter of the quantum dots, we choose the loss function for BO $f(\boldsymbol{X})$ dependent primarily on the value of hopping $t$, which spans many orders of magnitude. Because of the variations in value of $t$, the BO loss function must include the logarithm $\log_{10}(t(\boldsymbol{X})/\tilde{t})$. However, as $t\ll\tilde{t}$ in most of the voltage space, these regions need to correspond to a slowly varying loss function, and fast loss variations should only be allowed close to the minimum to guide the acquisition of new points. Therefore we define the BO loss function as follows:
\begin{equation}
f(\boldsymbol{X})=
\begin{cases}
\Bigg|\log_{10}\bigg(\frac{\big| \frac{t}{\tilde{t}}+1\big|}{2}\bigg)\Bigg|, & \text{SVM true}\\
C,& \text{SVM false}.
\end{cases}\label{loss}
\end{equation}
Function $f$ is shown in Fig. \ref{fig4a} in contrast to $\log_{10}(t(\boldsymbol{X})/\tilde{t})$, which invovles large value changes in unimportant regions of $t\ll\tilde{t}$. In Eq. \ref{loss}, $C$ is a constant value assigned to a case when the potential is rejected by SVM and it is introduced in order to constrain the optimisation domain of $\boldsymbol{X}$. We choose $C=\log_{10}\big(\frac{1}{2}\big)$, a value that $f(\boldsymbol{X})$ reaches in extreme case if $t$ is very small. An optimal combination $f(\boldsymbol{X}_{opt})$ for achieving desired $U$ as well as $t$ is selected from these sampled points which optimise well for $t$ but correspond to best $U$ as well.

\begin{figure}[H]
    \centering
    \includegraphics[trim=1.2cm 0 0 0,width=0.8\linewidth]{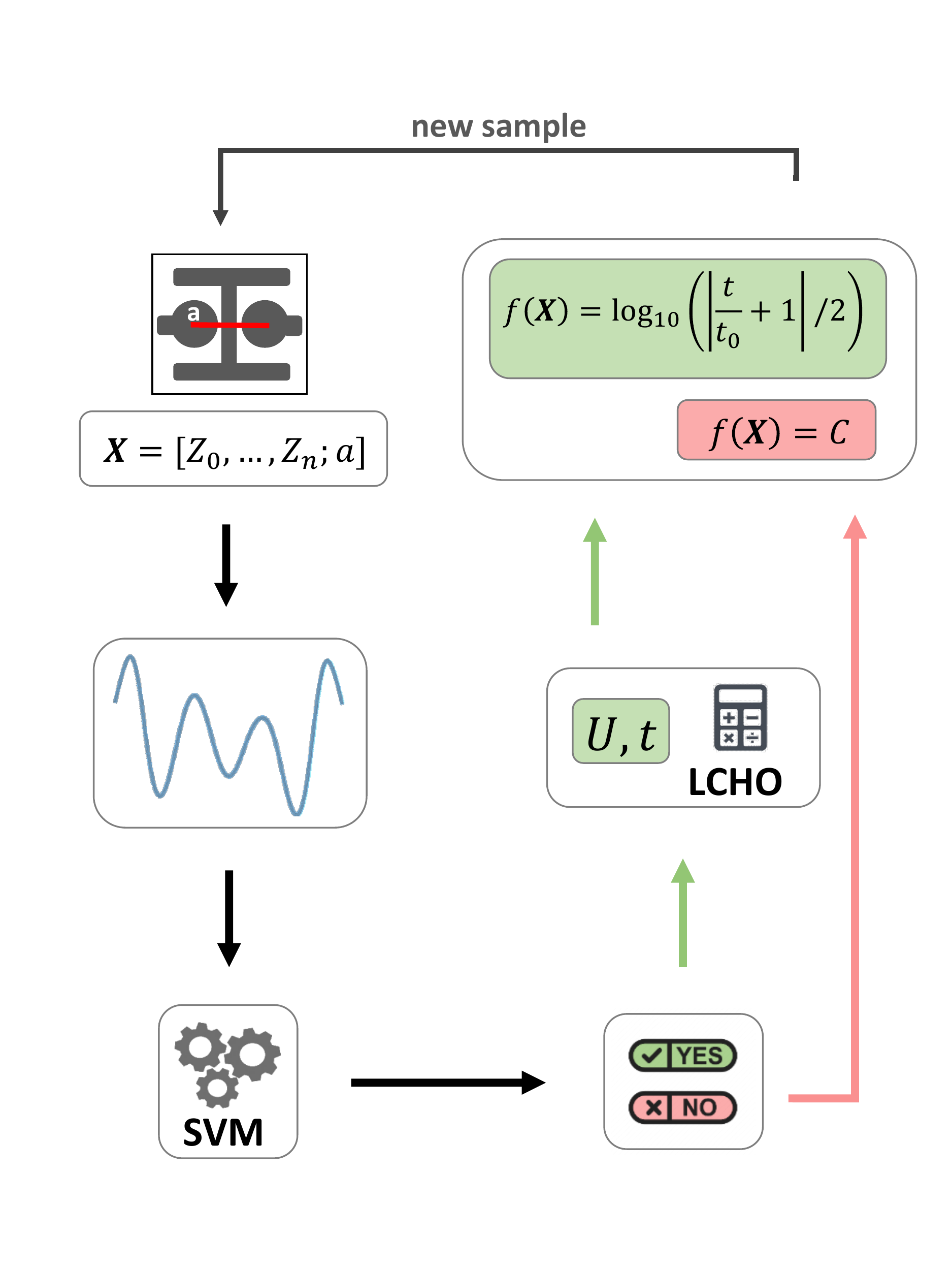}
    \caption{ BO loop for identification of optimal voltages for a desired Hubbard model with goal $t_0, U_0$. In each BO step a sample set of voltages and the QD distance $\boldsymbol{X}$ is selected for a fixed gate design (top left) and the electrostatic potential is obtained (middle left). Sample $\boldsymbol{X}$ is classified by the SVM (bottom left) as accepted (YES, bottom right) or rejected (NO, bottom right) based on the TB criterium. If $\boldsymbol{X}$ is accepted (green arrows), $t,U$ integrals are calculated with the LCHO method (middle right) and $t$ value is used to obtain the loss of $\boldsymbol{X}$, $f(\boldsymbol{X})$ (top right, green), otherwise (red arrow) the loss is constant $f(\boldsymbol{X})=C$ (top right, red). The pair $\boldsymbol{X},f(\boldsymbol{X})$ is added to other observed points $\{ \boldsymbol{X},f(\boldsymbol{X}) \}$, which are then used to select a new sample $\boldsymbol{X'}$.}
    \label{fig4}
\end{figure}

\begin{figure}[H]
    \centering
    \includegraphics[trim=0cm 0 0 0,width=0.97\linewidth]{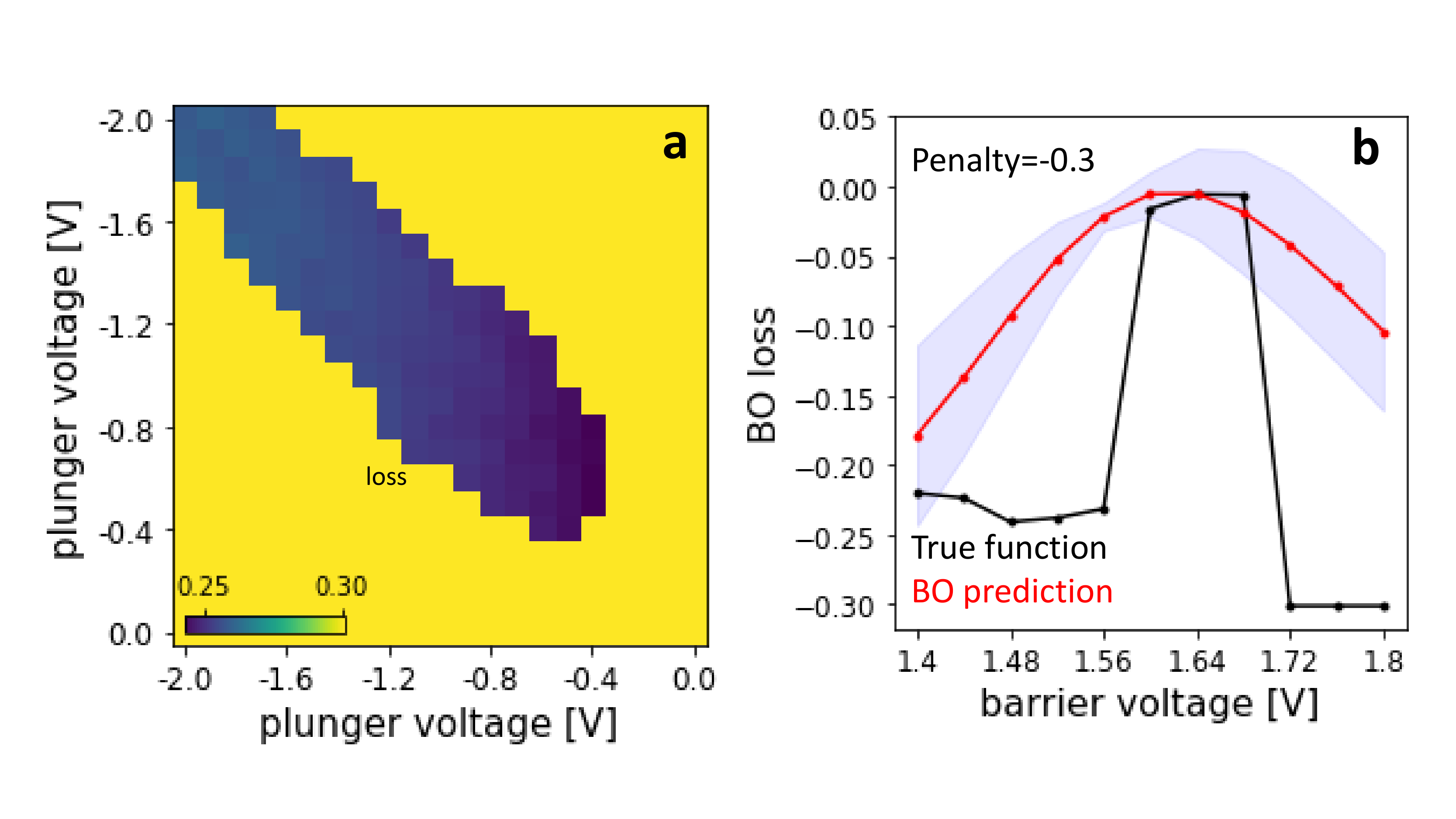}
    \caption{a) Loss function value $f(\boldsymbol{X})$ for a subspace of voltages along a plunger-plunger cut. Yellow (blue) denotes rejected (accepted) combinations. The darkest blue region identifies the minimum. b) Negative of the loss function $f(\boldsymbol{X})$ caculated (black) and predicted by the gaussian process within BO (red, variance in blue) for a linear space cut along the barrier voltage, close to the found minimum after 1900 steps. Gaussian process captures the overall trend of the loss, and agrees in value close to the minimum, where the combinations will be sampled.}
    \label{fig5}
\end{figure}

Fig. \ref{fig5} a) shows the value of the loss function in Eq. \ref{loss} for a cut in the voltage space (darker blue corresponds to lower loss values). Fig. b) shows the prediction of the underlying GP after $1900$ iterations for a single voltage path around a local optimum for $\boldsymbol{X}=(-0.84, -1.05, -0.29, 1.6, -0.09)$. The true loss values (GP mean) are shown in black (red) and the GP variance is shown in shaded blue. GP follows the trend of the true loss function, and agrees with it around the local maximum. This allows for efficient selection of new points to sample using the acquisition function.

\begin{figure}[H]
    \centering
    \includegraphics[trim=0cm 0 0 0,width=0.97\linewidth]{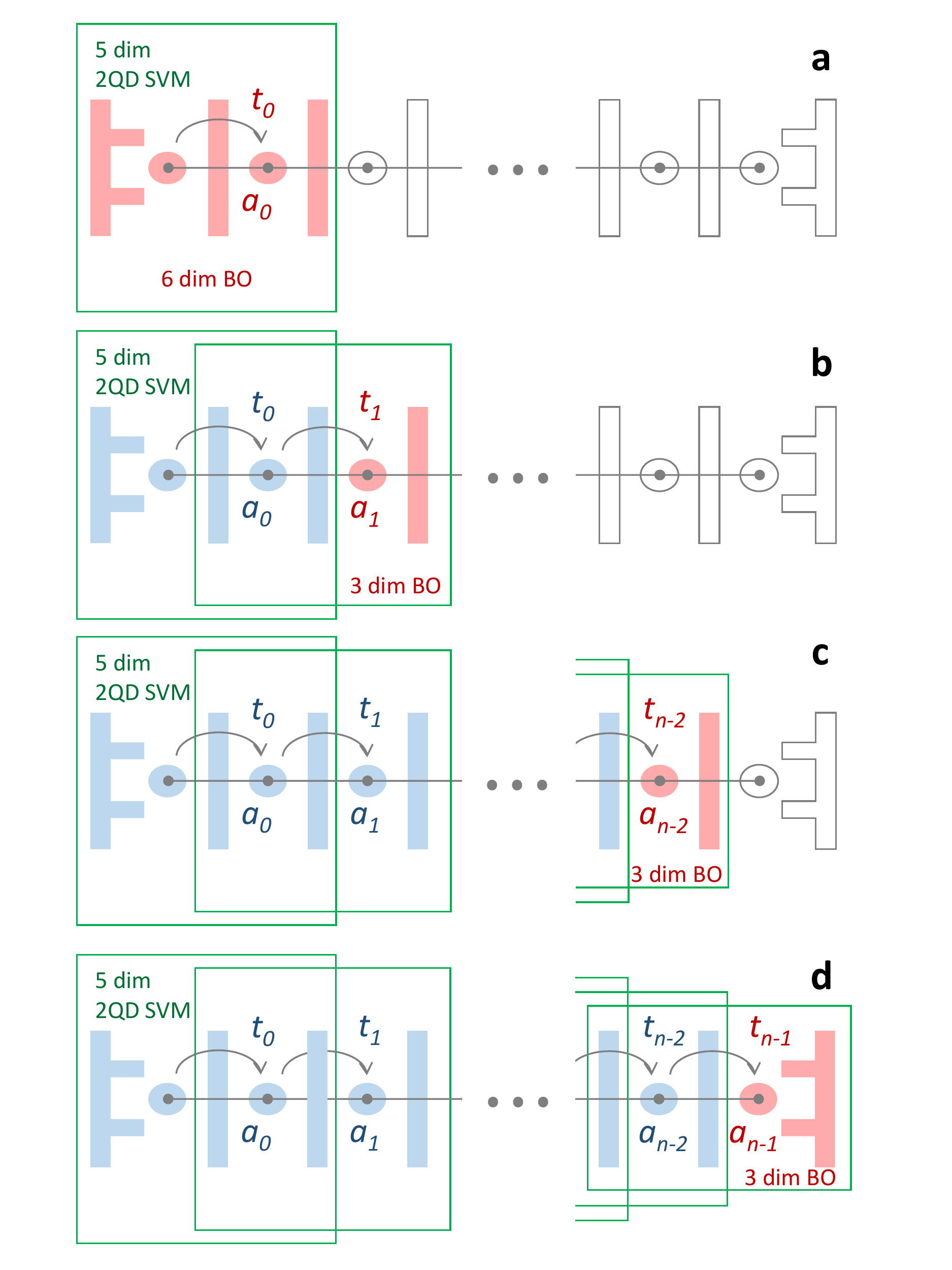}
    \caption{Iterative BO approach. a) Step 0: Using two-QD SVM (green), first 5 voltages (red) and the first 2-QD distance are optimised (as in Fig. \ref{fig4}). b) Step 1: Result $\boldsymbol{X}_0$ from step 0 is frozen (blue) and the next two voltages and the next QD distance are optimised (red). The same SVM (green) classifies a combination of frozen (blue) and variable (red) voltages. c) (d) Step N-2 (N-1) for N QDs, analogical to step 1, with all preceding results frozen.}
    \label{fig7}
\end{figure}

We now discuss the iterative approach to the voltage optimisation of linear arrays of more QDs, outlined in Fig. \ref{fig7}. We found that classifying subsets of voltages yields physical potentials, even for more than two QDs. We therefore use the unchanged SVM model built for two QD array to classify sets of five voltages a time, as shown with green boxes in Fig. \ref{fig7}. We begin with 6D BO optimisation of first five voltages (and lattice constant $a$), shown in red, and use the SVM model in each iteration for classification (Fig. \ref{fig7} a), shown in green. The outcome is the optimal $t_{opt}=t_{0}$ and the optimal parameters $\boldsymbol{X}_{opt}=(Z_0^{opt},Z_1^{opt},Z_2^{opt},Z_3^{opt},Z_4^{opt},a_0^{opt})$. In the next step, shown in Fig. \ref{fig7} b), we freeze the parameters obtained in the previous step (shown in blue), and perform a 3D BO of the pair of neighbouring voltages and distance to the next QD (shown in red). In this step, we still use the unaltered SVM model to classify every voltage sample, made of three frozen (blue) voltages and two variable (red) voltages. As a result we obtain next set of optimal parameters $\boldsymbol{X}_{opt}=(Z_5^{opt},Z_6^{opt},a_1^{opt})$ and the optimal $t_1$. We continue in this fashion (Fig. \ref{fig7} c), until the full structure has been optimised (Fig \ref{fig7} d).

The benefit of this iterative approach is that we are able to use BO for large systems without increasing the otpimisation problem dimension. Moreover, a single SVM model built for a small system can be used to optimise big structures. We also find that even for more QDs, a similar number of BO iterations is needed to achieve a satisfactory optimum. To ensure that all tunnelling integrals are within the range of the set goal we always pick the best optimum from a range of best solutions. This is because the neighbouring gates influence the previously optimised values, e.g. for three QDs, we found that the variation of $t_0$ during optimisation targeted for $t_1$ is approx. $\pm 25\%$ for all satisfactory $t_1$. We expect that this effect is local and will only significantly affect the neighbouring QDs in a bigger structure. To remedy the variations of the previously optimised $t$, we pick the best solution that achieves the goal for both $t$ values. This approach is straightforward in implementation, scalable to large systems and avoids the complexity of multi-targeted optimisation.

\section{ Results}

Here we present the results of our optimisation of gate voltages with a chosen goal of $\tilde{t}=0.8$meV for model gates and $\tilde{t}=0.2$meV for experimental gates. We run  BO for the 5 dimensional model gate voltage space with boundaries at $\boldsymbol{X}_{\text{min}}=[-2,-2,-2,-2,-2]$eV and $\boldsymbol{X}_{\text{max}}=[2,2,2,2,2]$eV. We test several $\lambda$  parameters for the $\text{UCB}_{\lambda}(\boldsymbol{X})$ function and choose the best results.

\begin{figure}[H]
    \centering
    \includegraphics[trim=3cm 0 3cm 0,width=0.95\linewidth]{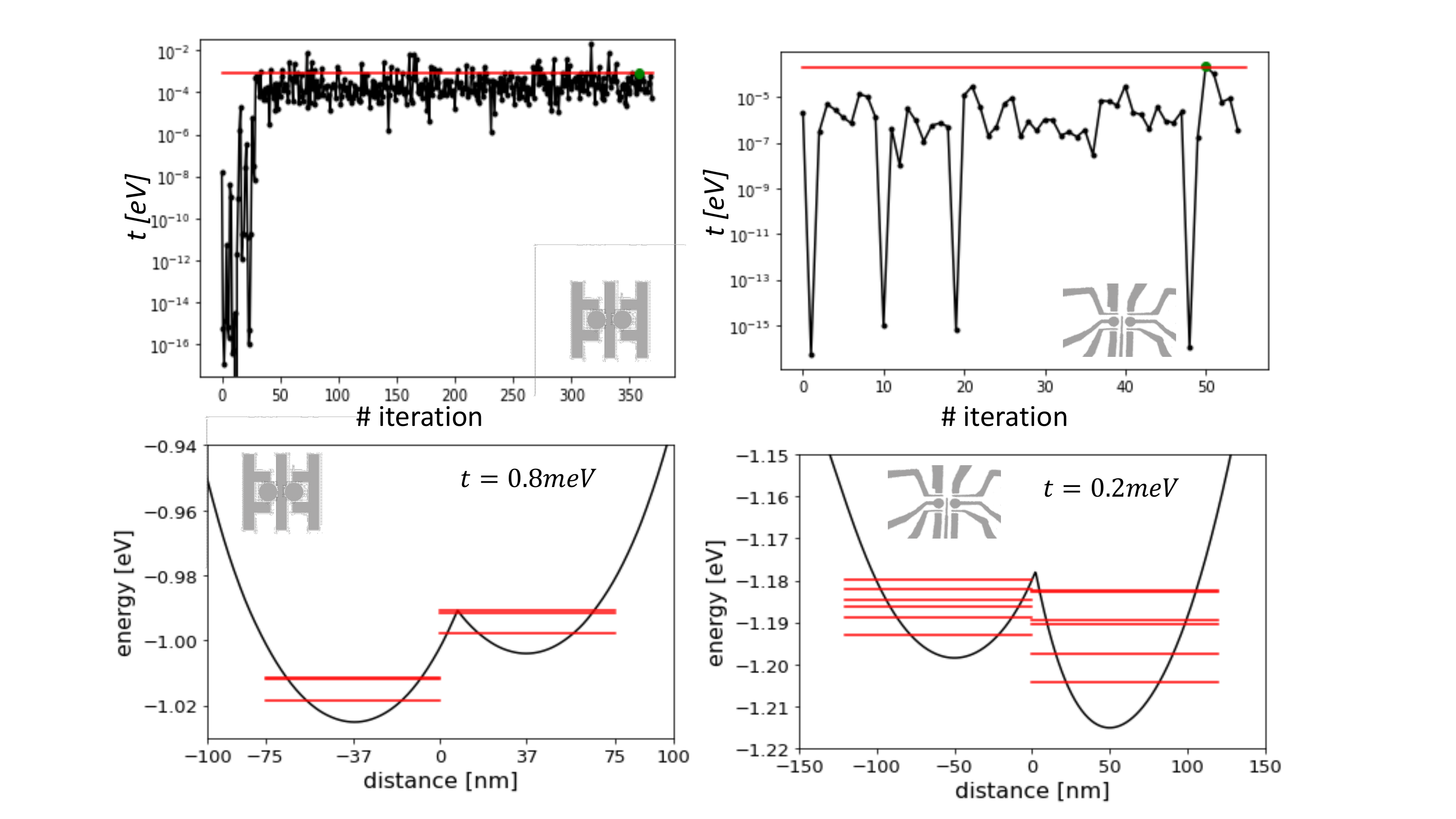}
    \caption{ a) (b) Values of $t$ obtained through BO as a function of iteration number for model (experimental) design. Red line marks the goal $\tilde{t}$ and green dot denotes the selected best combination. For model design, $t$ quickly approaches the goal and closes in on it as iterations progress. For experimental design fewer misses far away from goals are visible with time. c) (d) Resulting 2QD potentials corresponding to the green dots in a) (b) for model (experimental) designs. Red lines mark several first single particle energy levels for each QD to be occupied by particles in Hubbard model. For model design the QD potential bottom mismatch has not been restricted. The kink in the potentials is a consequence of the potential approximation for smaller QD distances.}
    \label{fig6}
\end{figure}

Fig. \ref{fig6} a) and b) show the loss values as the optimisation progresses for model and experimental gate designs respectively. The target $\tilde{t}$ has been marked with red line and the best solutions have been marked with green dots. Only iterations with positive SVM label are shown. It is apparent for both cases that the algorithm quickly abandons the regions of negligible $t$ and zooms in around local minima, while $t$ values grow. Initially around $10\%$ of new acquired points are labeled as positive and this proportion grows to about $50\%$.

Fig. \ref{fig6} c) and d) show the approximate potential for best found voltage and distance set $\boldsymbol{X}_{opt}=(Z_0^{opt},Z_1^{opt},...,a_{opt})$ produced by translation of single-QD potentials to optimal $a_{opt}$. The goal $\tilde{t}$ has been reached with $1.5\%$ error ($6\%$) for model (experimental) gates. Single-QD eigenenergies are shown in red in each quantum well. As no condition on QD bottom energy difference has been imposed, the bottom energies are $\approx20$meV apart in both cases. In order to achieve high hopping values $\tilde{t}$, a small distance $a_{lim}$ was needed. This calculation also served as a guide on possible values achievable for given designs. The resulting $a_{lim}\approx100$nm for experimental design is possible to reproduce in experiment with current fabrication technology \cite{hendrickx_four-qubit_2021, sajadi_2022}.

\begin{figure}[H]
    \centering
    \includegraphics[trim=3cm 0 3cm 0,width=0.7\linewidth]{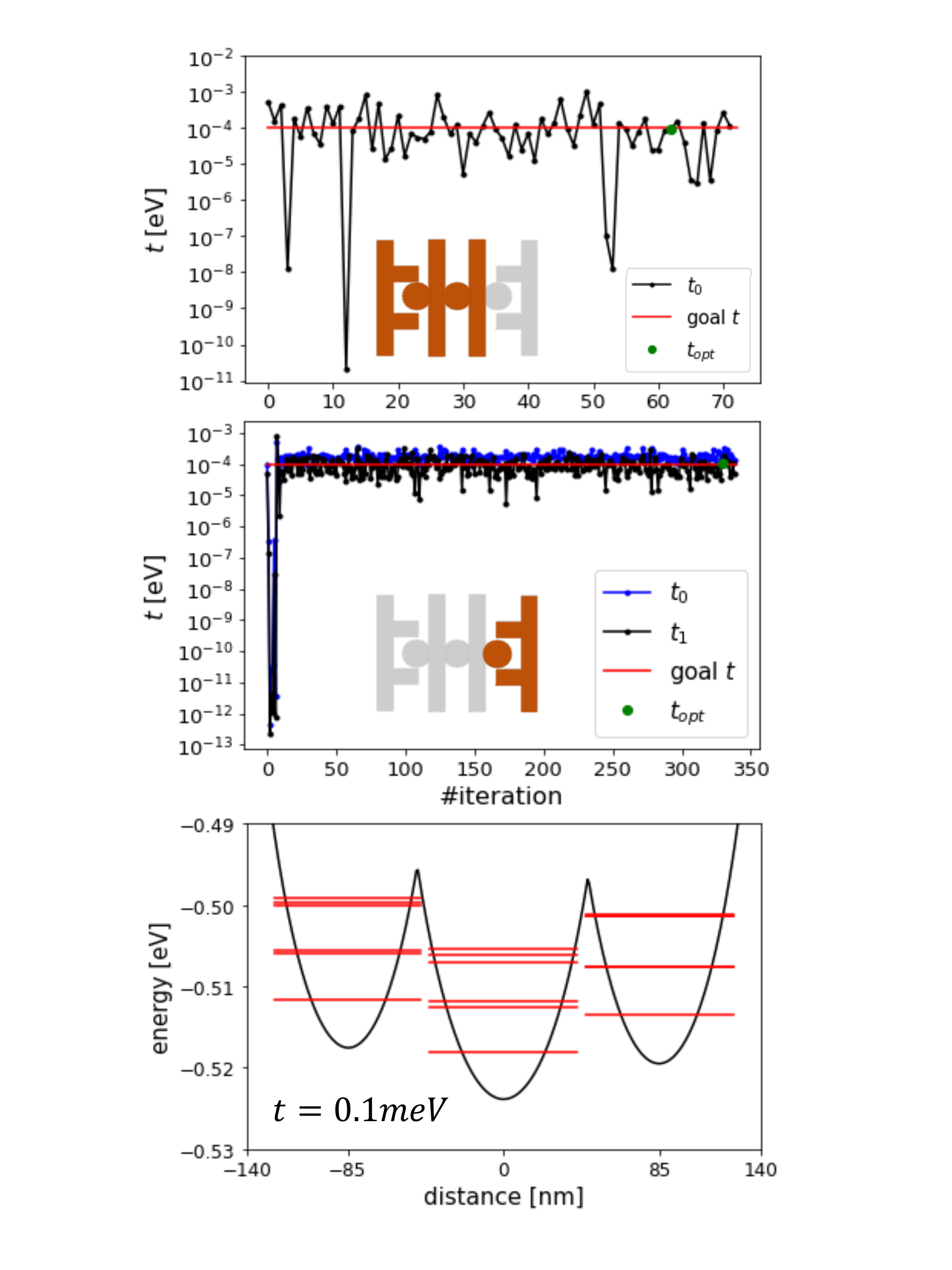}
    \caption{a) (b) Evolution of $t$ in BO loop (only accepted samples shown) for step 0 (1) of the iterative procedure for a three-QD array. Black denotes the value optimised in each step ($t_0$ in a), $t_1$ in b), blue is the $t_0$ value from preceding optimisation.  Red line denotes $\tilde{t}=0.1meV$. Red (grey) gates denote the optimised (frozen) voltages. c) Best result for 1000 samples in each step. Black (red) denotes the potential (SP energy levels). Kinks appear due to the approximation to potential at variable QD distance (see text).}
    \label{fig8}
\end{figure}

In Fig. \ref{fig8} we present the results of the iterative BO approach for three QD array. Fig. \ref{fig8} a) and b) show the evolution of the value of $t_0$ (black, goal shown in red) and $t_0, t_1$ (blue, black, goal shown in red) in steps $0$ and $1$ of the iterative procedure. Red gates are varied while grey remain frozen. Both steps included 1000 BO iterations, but due to higher dimensionality of step $0$ than step $1$, fewer points in space are accepted by the SVM model, so the curve is more sparse. Fig. \ref{fig8} b) shows that for satisfactory $t_1$, variations of $t_0$ are much smaller than $t_1$, which allows us to pick pairs of $t$ best meeting the goal.

Fig. \ref{fig8} c) shows the best three QD potentials that achieves both goals of $\tilde{t}=0.1meV$ within $10\%$ error, while maintaining the alignment of the QD SP energy level ranges. The optimised distance for both steps $0$ and $1$ was found to be the same, up to less than $1\%$.

\section{ Conclusions}
    
We used machine learning techniques for the experimental voltage setup optimization that produces the target Hubbard $t$ and $U$ parameters and allows for tailored Hubbard model simulation. Using SVM we classified potential profiles based on suitability for tight-binding approximation reducing the space by 99\%. We then optimized the voltage combinations using Bayesian optimization, which operates without evaluating gradients of the loss function or experimental stability diagram input. Our results rely on accurate experimental gate images and realistic integrals based on LCHO calculation. We were able to predict voltages needed in experiment to prepare on demand double QD Hubbard model within 1.5\% error for model gates and 6\% error for experimental gates. 

We also designed an iterative procedure using several Bayesian optimisation instances, one for each extra QD above two QDs. It involves the optimisation of subsets of voltages, which allows for scaling to large systems, and targets a single tunnelling integral at a time, which avoids high complexity of multi-target approaches. Importantly, we use the same SVM model trained for two QDs to classify subsets of voltages in each iteration, so training exponentially hard SVM models can be avoided for bigger arrays. We reach an accuracy of 10\% for both tunnelling parameters $t$ within the same number of iterations as used for two-QD system.

This procedure demonstrates how on-demand Hubbard models can be prepared in experiments to explore new Hubbard physics. It could be embedded in a larger algorithm to reliably tune tunnelling couplings with BO and prepare charge states using existing experiment-based methods.

\section{Acknowledgements}
We thank E. Sajadi, M. Tanvir and J. Salfi for discussions and permission to use the experimental image in Fig. 1. b).

\bibliography{introBib}


\end{document}